\documentclass[aps,pre,superscriptaddress,twocolumn,amsmath,amssymb,showpacs]{revtex4}
\usepackage{graphicx}
\usepackage{dcolumn}
\usepackage{bm}
\usepackage{textcomp}

\begin{document}

\title{Optimal search 
in interacting populations:
Gaussian jumps vs L\'evy flights}

\author{Ricardo Mart\'inez-Garc\'ia}
\affiliation{IFISC, Instituto de F\'isica Interdisciplinar y Sistemas Complejos (CSIC-UIB),
  E-07122 Palma de Mallorca, Spain}

\author{Justin M. Calabrese}
\affiliation{Conservation Ecology Center,
Smithsonian Conservation Biology Institute,
National Zoological Park,
1500 Remount Rd., Front Royal, VA 22630, USA}

\author{Crist\'obal L\'opez} 
\affiliation{IFISC, Instituto de F\'isica Interdisciplinar y Sistemas Complejos (CSIC-UIB),
  E-07122 Palma de Mallorca, Spain}

\date{\today}
\begin{abstract}
We investigated the relationships between search efficiency, movement strategy, and non-local communication in the biological context of animal foraging. We considered situations where the members of a population of foragers perform either Gaussian jumps or L\'evy flights, and
show that the search time is minimized when communication among individuals occurs at intermediate ranges, independently
of the type of movement. Additionally, while Brownian strategies are more strongly influenced by the communication mechanism, 
L\'evy flights still result in shorter overall search durations.
\end{abstract}


\maketitle
 \section{Introduction}\label{introsec}

Situations where a single individual or a group of searchers must find an object (target)
appear in many different fields including chemistry \cite{rmp_benichou,mendez2014random},
information theory \cite{pirolli},
and animal foraging \cite{libroforaging}. The study of these searching problems
has generated an increasing amount of work in the last years, many of them
oriented towards the identification of efficient strategies  
\cite{rmp_benichou,libroforaging,nature-vergassola,benPRE2006}.
Many remarkable examples can be found in the context of biological
encounters,
such as proteins searching for targets on DNA \cite{bio-taylor}, or animals searching
for a mate, shelter or food 
\cite{Campos2013,rmp_benichou,visna1999,shlesinger2006,edwards2007, Torney2011, hein, Viswanathan2008, mejiamonasterio,bartomeus}. 
In these cases, the search time is generally limiting and minimizing it
can increase individual fitness or reaction rates.

The optimality of a search strategy depends 
strongly on the nature of both the targets and the searchers
\cite{bartumeus2005animal,BartumeusPRL}. In the context of animal foraging, which is our focus here, searchers 
may move randomly, may use memory and experience to locate dispersed targets or they may also combine random search with memory-based search.
In highly social species, groups of searchers may share 
information when no single individual is sufficiently 
knowledgeable. 
This is based on the many wrong 
hypothesis \cite{hoare, torneypnas}, that states that error in sensing of individuals
can be reduced by interacting with the rest of the group, where all individuals can act as a
sensors.

It is well known that individual movement plays a central role in search efficiency, and many studies have focused
on the comparative efficiency of L\'evy and  
Brownian movement strategies \cite{rmp_benichou,Viswanathan2008,bartumeus2005animal,BartumeusPRL}.
L\'evy flights are more efficient in some random search scenarios \cite{visna1999,viswaphysica}, but whether or not they
are used in real animal search strategies is still an open and contentious topic \cite{edwards2011overturning,edwards2007}.
Much less effort, however,
has been spent on trying to understand the long-range (i.e. nonlocal) interaction
mechanisms among social searchers. While diverse
observations suggest that such interactions occur in many taxa, including 
bacteria \cite{LiuPassino}, insects, and mammals \cite{dianamonkey,mccomb},
previous studies have focused almost exclusively on how the collective movements of a group of animals can emerge from
local interactions among individuals \cite{PhysRevE.86.011901,Kolpas2013,Couzin2002}.
To our knowledge, only two recent studies have explored the effects of long-range communication
mechanisms on a searching strategy \cite{Torney2011,Martinez-Garcia2013b}.
In particular, \cite{Martinez-Garcia2013b} showed that when the communication range is intermediate,
individuals tend to receive the optimal amount of information on the locations of targets, and 
search time is consequently minimized.
Longer communication ranges overwhelm the searchers as they are simultaneously called from all directions, while shorter ranges 
do not provide enough information. However, many open questions
remain about the relationship between communication
and search efficiency,
especially concerning the role that the landscape plays
in determining the optimal communication range, and on the 
robustness of the behavior of the model when different random movements are considered. 
Here, we compare the effects of non-local communication on the search efficiency of groups of individuals employing either L\'evy flight or Brownian random search strategies.
We also investigate how the distance between targets influences the optimal communication range for both strategies. 
For tractability, we consider a simplified, one-dimensional version of 
the model and compute analytically the search time for both
Brownian and L\'evy flights as a function of the communication length scale. This simplified model
allows us to unveil the dependence of 
search time on both the parameters governing individual mobility, and on 
the distance between targets. 

The paper is organized as follows. The general model is presented in Sec.~\ref{modelsec}.
Sec.~\ref{brownsec} and Sec.~\ref{levysec} presents analytical and numerical results
for Brownian and L\'evy strategies, respectively.
In Sec.~\ref{comparation}, the role of the communication mechanisms in the different
searching strategies is discussed, and the paper ends with Sec.~\ref{summmary}, where a summary and
conclusions are presented.

\section{The model}\label{modelsec}

{\it General.} 
We consider a population of $N$ interacting individuals that move randomly searching for spatially
distributed targets. Every individual is provided with information about the location of 
the targets coming from two different sources: its own perception ({\it local information}), 
and the knowledge on the quality of far away areas coming 
from a communication mechanism with the rest of the population ({\it nonlocal information}).
The Langevin equation describing this dynamics is,
\begin{equation}\label{lange}
 \dot{\mathbf{r}_i}(t)=B_{g}\nabla g(\mathbf{r}_i)+B_{C}\nabla S(\mathbf{r}_i)+\mathbf{\eta}_{i}(t), \ \ \ \ i=1, \ ... \ N.
\end{equation}
where $\mathbf{\eta}_{i}(t)$ is a stochastic function of time, which we specify below, whose statistical properties
define the random motion performed by the searchers.
 The term $B_{g}\nabla g(\mathbf{r}_i)$ refers to the local search, where $g(\mathbf{r})$ is an
environmental quality function (amount of grass, prey, etc...). $g$ takes values between
$0$ (low quality areas) and $1$ (high quality areas), and thus allows us to define the targets of the
search as those regions where the environmental quality is higher than a given
threshold, $\kappa$.
$B_{g}$ is the local search bias parameter. $B_{C}\nabla S(\mathbf{r}_i)$ is the nonlocal 
search term, where $B_{C}$ is the nonlocal search bias parameter and $S(\mathbf{r}_i)$
is the {\it nonlocal information function} of the particle $i$. 
It represents the information arriving at the position of the individual $i$ as a result
of communication with the rest of the population. The net effect of these two terms is to drift the movement
of the searcher towards high quality areas of the environment. The model thus becomes an Ornstein-Uhlenbeck process,
with individuals drifting randomly, but with an attraction to the location of the targets \cite{smouse, Preisler, Turchin}.

Following previous efforts \cite{Martinez-Garcia2013b, Dean}, the interaction among individuals is given in terms of a  
nonlocal function defined as the superposition of the pairwise interaction between one
individual and each one of the other members of the population,
\begin{equation}\label{colective}
S(\mathbf{r}_i)=\left(\sum_{j=1, j\neq i}^{N} A[g(\mathbf{r}_j)]V(\mathbf{r_i},\mathbf{r}_j)\right),
\end{equation}
where $V(\mathbf{r}_i,\mathbf{r}_j)$ is the two-body interaction between the receptor, $i$, at
$\mathbf{r}_i$ and the emitting particle fixed at $\mathbf{r}_j$. $A[g(\mathbf{r}_j)]$ is an 
activation function (typically, a Heaviside function) that turns on when the individual
at $\mathbf{r}_j$ has found a target and starts communicating.

\begin{figure}
\begin{center} 
\includegraphics[width=.4\textwidth]{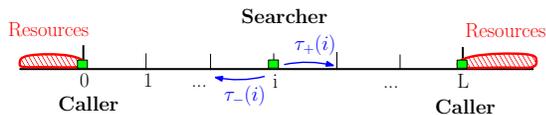}
\caption{(Color online). Scheme of the simple version of the model.}
\label{scheme}
\end{center}
\end{figure}

A study of the behavior of this model in $2D$ based on Monte Carlo simulations,
and using a Gaussian white noise for the function $\eta(t)$, i.e. Brownian motion,
has been shown in \cite{Martinez-Garcia2013b}. To gain clearer insight and provide analytical 
arguments, we study a minimalistic version of the model.

{\it Specific considerations.}
Consider a single individual in a one-dimensional space of length $L$, so
that the highest quality areas are located beyond the limits of the 
system, i.e. at $x=-1$ and $x=L+1$ (see Fig.~\ref{scheme}).
Note that this would correspond to the ideal situation where all
the members of the population but one -the searcher- have
already reached one of the targets. 
A landscape quality function, $g(x)$ must also be defined.
Provided it is a smooth, well-behaved function, its particular shape 
is not relevant. We therefore assume a Gaussian-like quality landscape,
\begin{equation}\label{qlandscape}
g(x)={\rm e}^{-\frac{(x+1)^{2}}{\sigma_{r}}}+{\rm e}^{-\frac{(x-L-1)^{2}}{\sigma_{r}}},
\end{equation}
where $\sigma_{r}$ gives the characteristic width of a high quality region.
Notice that $g(x)$ is defined so that 
highest quality areas are located, as mentioned,  at $x=-1$ and $x=L+1$.
This ensures that the gradient of the function does not vanish at the extremes of the system
(Figure \ref{scheme}), and it is equivalent to setting the value of the threshold $\kappa$ such that the targets
start at $x=0$ and $x=L$. We assume that a foraging area is good enough when its quality 
is higher than $80\%$ of the ideal environment, which means $\kappa=0.8$. As we center the patches of resources
at $x=-1$ and $x=L+1$, fixing a good quality treshold at $\kappa=0.8$ is equivalent to fix the width of the environmental quality function at $\sigma_{r}=4.5$,
to ensure that $g(0)=g(L)\approx0.80$. However, the qualitative behavior of the model is independent of this choice.

Finally, the pairwise communication function needs to be specified, and we choose a family of functions given by 
\begin{equation}\label{twobody}
V(x_{i},x_{j})=\exp\left(-\frac{|x_{i}-x_{j}|^{p}}{\sigma}\right),
\end{equation}
 where $\sigma^{1/p}$ gives the
typical communication scale. For simplicity, and without loss of generality, we will consider only the case $p=2$.
Indeed, the choice of the function $V$ is not relevant for the behavior of the model, provided that
it defines an interaction length scale through the parameter $\sigma$. This scale
must tend to zero in the limit $\sigma\rightarrow 0$ and to
infinity in the limit $\sigma\rightarrow \infty$. This assures that the 
gradient of the calling function vanishes in these limits. 
The combination of local and nonlocal information gives the total available information for the searcher,
$R(x)=B_{g}g(x)+B_{c}S(x)$. 

To obtain analytical results, we work in the following on a discrete space.
The stochastic particle dynamics equivalent to  Eq.~(\ref{lange}) considers 
left and right jumping rates which are  defined for every individual using the total information function,
\begin{equation}\label{rates}
 \tau_{\pm}(x)=\max\left(\tau_{0}+\frac{R(x\pm h)-R(x)}{h},\alpha\right),
\end{equation}
where $\alpha$ is a small positive constant to avoid negative rates that has been given 
an arbitrary value $(\alpha=10^{-4})$, and $h$ is the spatial discretization ($h=1$).
Finally, $\tau_{0}$ is the jumping rate of an individual in absence of information, and it is related 
to the diffusion component of the dynamics of Eq.~(\ref{lange}).
Given the transition rates of Eq.~(\ref{rates}), the movement with a higher gain of information has a higher rate, and therefore a larger probability of taking 
place. 

The simplest situation, which allows an analytical treatment of the problem, is to consider only $N=3$ individuals. Two of them
are located in the top quality areas just beyond the frontiers of the system limit, $x=-1$ and $x=L+1$,
and the other one is still searching for a target. 
Under these considerations, using the environmental quality function defined in Eq.~(\ref{qlandscape}), and the 
pairwise potential of Eq.~(\ref{twobody}), the total available information for the searcher is
\begin{eqnarray}\label{totalinfo}
 R(x;\sigma,L)=B_{g}\left({\rm e}^{-\frac{(x+1)^{2}}{\sigma_{r}}}+{\rm e}^{-\frac{(x-L-1)^{2}}
{\sigma_{r}}}\right) \nonumber \\
+B_{C}\left({\rm e}^{-\frac{(x-L-1)^{2}}{\sigma}}+{\rm e}^{-\frac{(x+1)^{2}}{\sigma}}\right).
\end{eqnarray}

Following \cite{Martinez-Garcia2013b}, the efficiency of the search process is measured in terms of the first arrival
time at one of the high quality areas, either at $x=0$ or $x=L$,  starting from $x_{0}=L/2$. From the definition of the transition
rates in Eq.~(\ref{rates}), $\tau_{+}(L-1)\gg\tau_{-}(L)$,
and equivalently $\tau_{-}(1)\gg\tau_{+}(0)$. This means that at both extremes of the system, the 
rate at which particles arrive is much higher than the rate at which they leave, so particles do not move
when they arrive in the top quality areas. This allows us to consider  
both extremes $x=0$ and $x=L$ of the system as absorbing, 
and the first arrival time may be obtained from the flux of presence probability of the searcher there \cite{redner}
\begin{equation}\label{meantime}
 \langle T(\sigma)\rangle=\int_{0}^{\infty}t\left(\frac{\partial P(0,t)}{\partial t}+\frac{\partial P(L,t)}{\partial t}\right)dt.
\end{equation}
This definition will be used in the following sections to investigate
the influence of sharing information (i.e., of the interaction mechanism) on search times. The results will be compared with
those obtained using a deterministic approximation of the movement of the searcher.
We study two different random strategies--Brownian and L\'evy.

\section{Brownian jumps}\label{brownsec}

\begin{figure}
\begin{center} 
\centering
\includegraphics[scale=0.35]{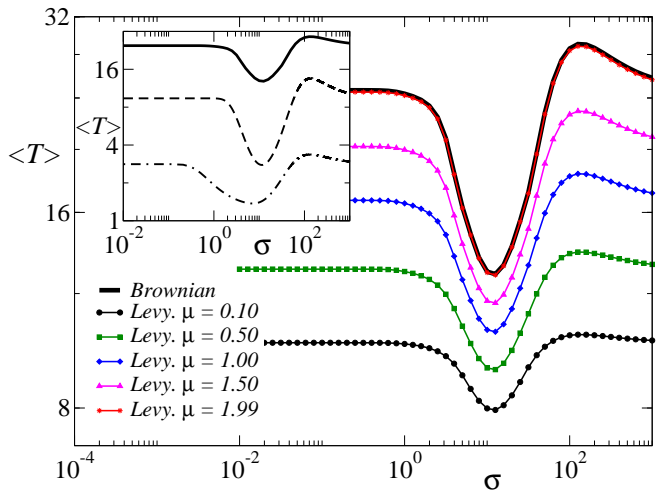}
\caption{(Color online). First arrival time solving Eq.~(\ref{diffset}) for the Brownian jumps and Eq.~(\ref{diffsetlevy}) in the case of L\'evy flights for different values of $\mu$. Lines are 
interpolations. Inset: First arrival time using its definition Eq.~(\ref{meantime}) (full line) and Eq.~(\ref{time}) with $\epsilon=2$ (dashed line) and $\epsilon=0$ (dotted dashed line) for a Brownian searcher. 
In both panels: $L=9$, $\sigma_{r}=4.5$, $B_{g}=1$, and $B_{c}=1$.}
\label{timesana}
\end{center}
\end{figure}

 In this case the searcher only
jumps -with a given rate- to its nearest neighbors, so the coupling of the set of differential equations describing the occupancy probability 
of every site of the system is (notice that lattice spacing $h=1$),
\begin{eqnarray}\label{diffset}
 \frac{\partial P(0,t)}{\partial t}&=&-\tau_{+}(0)P(0,t)+\tau_{-}(1)P(1,t), \nonumber \\
 \frac{\partial P(i,t)}{\partial t}&=&-(\tau_{+}(i)+\tau_{-}(i))P(i,t)+ \nonumber \\
&&\tau_{+}(i-1)P(i-1,t)+\tau_{-}(i+1)P(i+1,t), \nonumber \\
 \frac{\partial P(L,t)}{\partial t}&=&-\tau_{-}(L)P(L,t)+\tau_{+}(L-1)P(L-1,t).
\end{eqnarray}
with $i=1,...,L-1$. 
If the initial position of the particle is known, it
is possible to solve Eq.~(\ref{diffset}) using the Laplace transform.
Once the probability distribution of each point has been obtained, it is possible to obtain
the mean first arrival time using Eq.~(\ref{meantime}).
The thick line in Fig.~\ref{timesana} shows this result, indicating that the searching process is optimal
(minimal time to arrive to one of the good quality areas) 
for intermediate values of $\sigma$.
A particularly simple limit in Eq.~(\ref{diffset}) appears when 
$\tau_{+}>>\tau_{-}$ when $x>L/2$ (and the contrary on the 
other half of the system). The search time is  $T(\sigma)=\frac{L}{2\tau_{+}}$.
This is the expected result since the movement is mainly in one direction
and at a constant rate.

In biological terms this means 
that the optimal situation for the individuals is to deal with intermediate amounts of information.
Extreme situations, where too much ($\sigma\rightarrow\infty$) or too little ($\sigma\rightarrow0$) information is provided by the population,
have the same effect on the mean first arrival time, which tends to the same asymptotic value in both limits.
In both cases, the search is driven only by the local perception of the environment \cite{Martinez-Garcia2013b}.

This calculation gives exact results, but it implies
fixing the system size, solving a set of equations of dimension $L$, and finally
obtaining the inverse Laplace transform of the solutions. The main disadvantage of this approach
is that it is not possible to study the influence of the distance
between targets on the optimal communication length. To circumvent this we
use a deterministic approach in 
the continuum  limit $h\rightarrow 0$ 
and define, using the symmetry of the system, 
a mean drift velocity towards one of the high quality areas, $x=L$,
\begin{equation}\label{average}
\langle v_{d}(\sigma, L)\rangle =\int_{L/2}^{L}(\tau_{+}(x)-\tau_{-}(x))dx,
\end{equation}
Substituting the definition of the transition rates Eq.~(\ref{rates}), the drift velocity is,
\begin{equation}\label{drift}
 \langle v_{d}(\sigma, L)\rangle=2\left[R\left(L\right)-R\left(\frac{L}{2}\right)\right],
\end{equation}
and therefore the search time is
\begin{equation}\label{time}
\langle T(\sigma,L)\rangle=\frac{N/2}{\langle v_{d}(\sigma,L)\rangle}.
\end{equation}

 We compute the searching time using Eq.~(\ref{time})
with the same values of the parameters used before ($\sigma_{r}=4.5$, $B_{g}=1$, and $B_{c}=1$, $L=9$) 
to compare it with the results given by Eq.~(\ref{meantime})
(inset of Figure \ref{timesana}). The approach in Eq.~(\ref{time}) (dotted-dashed line) reproduces the qualitative behavior
of the searching time although underestimates the value of the optimal communication range ($\sigma_{opt}=7.2$
while Eq.~(\ref{meantime}) produces $\sigma_{opt}=12.5$). This can be fixed excluding from the average in Eq.(\ref{average})
the boundary of the system introducing a parameter $\epsilon$ in the limits of the integration. Results for $\epsilon=2$ 
correspond to the dashed line in the inset of Figure \ref{timesana} (See Appendix \ref{app2} for details). However, regardless of 
the value of $\epsilon$ used in the average, this approximation underestimates the 
temporal scale of the problem (the absolute values of the times).
This is because it is assumed that the searcher follows a deterministic
movement to the target neglecting any fluctuation that may slow the process.

Finally, increasing $\sigma$ beyond its optimal value, 
there is a maximum for the search time for any of the approaches.
For these values of the communication range, the nonlocal information
at the middle of the system coming from both targets is higher than in the extremes and
thus there is a bias to the middle in the movement of the searcher.
This small effect, that vanishes when $\sigma$ increases and the information tends to be constant
in the whole system, seems to be an artifact of the particular arrangement 
of the simplified $1-D$ system, and does not seem relevant for any real-world consideration of this kind of model.
In addition, it does not substantially affect the dynamics because local perception of the environment
pushes the individual towards one of the targets.

\begin{figure}
\begin{center} 
\centering
\includegraphics[scale=0.30]{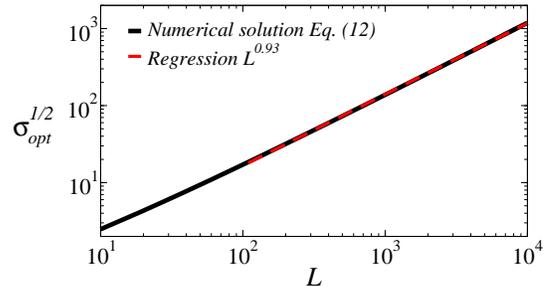}
\caption{(Color online). Scaling of the optimal communication range parameter with the distance between targets (system size in the $1-D$ simple model).}
\label{scaling}
\end{center}
\end{figure}

Finally, within this deterministic approximation, besides studying larger systems with no additional computational cost,
it is possible to obtain the optimal value of the interaction range parameter, $\sigma_{opt}$: 
\begin{equation}\label{optsigma}
\left(\frac{\partial T}{\partial \sigma}\right)_{\sigma=\sigma_{opt}}=0, 
\end{equation}
which has to be solved numerically for different sizes of the system. 
 The typical optimal communication scale defined by $\sigma^{1/p}$, (i.e., by $\sigma^{1/2}$ since $p=2$)
grows approximately linearly with the distance between targets in the asymptotic limit. Using a regression of the results obtained from the integration 
of Eq.~(\ref{optsigma}) yields an exponent $\sigma_{opt}^{1/2}\propto L^{0.93}$ for $L\gg1$ (Figure \ref{scaling}). 

\section{L\'evy flights}\label{levysec}

So far, the model has been studied assuming that the searchers employ Brownian motion. 
Alternatively, L\'evy flights have been shown as a good searching strategy that
may be used by some species. However, empirical studies have generated controversy, since many of 
the statistical methods used to support the presence of L\'evy flights in nature have been questioned,
and the issue remains unresolved \cite{nature_hum,edwards2007,edwards2011overturning}.
In this section the case of L\'evy searchers is considered. The results will show that neither the behavior of the model, 
nor the existence of an intermediate optimal communication scale, depend on the characteristics of the motion of the individuals.

L\'evy flights do not have a typical length scale and thus
the searcher can, in principle, make jumps as larger as the size of the system. The lengths of the jumps, $l>0$, are sorted from a 
probability distribution with a long tail \cite{metzler,klages}
\begin{equation}\label{levyasymp}
P_{\mu}(l)\approx \tilde{l}^{\mu}l^{-(\mu+1)}, \ \ l\rightarrow\infty,
\end{equation}
with $l\gg\tilde{l}$, and $0<\mu<2$, where $\tilde{l}$ is a characteristic length scale of the system.
This distribution is
not defined for $\mu<0$, its mean and variance are unbounded for $0<\mu\leq1$, and it has a mean but no variance for $1<\mu<2$. 
Finally, for $\mu\geq2$, the two first moments exist and thus it obeys the central limit theorem. The Brownian motion 
limit is recovered in this latter case, while very long jumps are more frequent in the ballistic limit, when $\mu\rightarrow 0$.
The cumulative distribution corresponding to Eq.~(\ref{levyasymp}) is 
\begin{equation}\label{cumnonor}
 \Psi_{\mu}(l)\approx\mu^{-1}\left(\frac{l}{\tilde{l}}\right)^{-\mu}, \ \ l\rightarrow\infty.
\end{equation}
As a simple normalizable cumulative distribution function, with the
asymptotic behavior of Eq.~(\ref{cumnonor}), we will use \cite{Elsref}
\begin{equation}\label{cumnor}
 \Psi_{\mu}(l)=\frac{1}{\tilde{l}\left(1+\frac{l}{\tilde{l}} b^{1/\mu}\right)^{\mu}},
\end{equation}
whose probability distribution, $P_{\mu}(l)=\Psi_{\mu}'(l)$, is given by
\begin{equation}\label{levydist}
 P_{\mu}(l)=\frac{\mu b^{1/\mu}}{\tilde{l}\left(1+\frac{l}{\tilde{l}} b^{1/\mu }\right)^{\mu+1}},
\end{equation}
with $0<\mu<2$, and $b=[\Gamma(1-\mu/2)\Gamma(\mu/2)]/\Gamma(\mu)$. 
We fix $\tilde{l}=h=1$, and the transition rate as defined in Eq.~(\ref{rates}).

{\it Results in one dimension.}
Proceeding similarly to the previous section, 
the set of equations for the probability of occupancy is 
\begin{widetext}
\begin{eqnarray}\label{diffsetlevy}
 \frac{\partial P(0,t)}{\partial t}&=&\sum_{j=1}^{L}\tau_{-}(j)B_{j}P(j,t)-\tau_{+}(0)P(0,t)\left(B_{L}+\sum_{j=1}^{L-1}A_{j}\right), \nonumber \\
  \frac{\partial P(i,t)}{\partial t}&=&\sum_{j=0}^{i-1}\tau_{+}(j)A_{i-j}P(j,t)+\sum_{j=i+1}^{L}\tau_{-}(j)A_{j-i}P(j,t)- \nonumber \\
  &&\tau_{-}(i)P(i,t)\left(B_{i}+\sum_{j=1}^{i-1}A_{j}\right)-\tau_{+}(i)P(i,t)\left(B_{L-i}+\sum_{j=1}^{L-i-1}A_{j}\right),  \ \ \ i=1,\ldots,L-1 \nonumber \\
 \frac{\partial P(L,t)}{\partial t}&=&\sum_{j=0}^{L-1}\tau_{+}(j)B_{L-j}P(j,t)-\tau_{-}(L)P(L,t)\left(B_{L}+\sum_{j=1}^{L-1}A_{j}\right). \nonumber \\
\end{eqnarray}
\end{widetext}

We assume that if a jump of length in between $j-1$ and $j$ takes place, the individual gets the position $j$.
To this aim, the coefficients $A_{j}$ enter in the set of equations Eq.~(\ref{diffsetlevy}) and are defined as $A_{j}=\int_{j-1}^{j}\Psi_{\mu}(l)dl$. They give the probability
of a jump of length between $j-1$ and $j$ to happen. The coefficients $B_{j}$ are defined as $B_{j}=\int_{j-1}^{\infty}\Psi_{\mu}(l)dl$, to 
take into account that the searcher stops if it arrives to a target. This introduces a cutoff in the jumping length distribution Eq.~(\ref{levydist}). 

Given the size of the system, $L$, which fixes the dimension of the system of equations Eq.~(\ref{diffsetlevy}), it is possible
to obtain an analytical solution for the occupancy probabilities and the mean arrival time to the
targets using Eq.~(\ref{meantime}).
This is shown in Figure \ref{timesana}, where the Brownian 
limit is recovered when $\mu\rightarrow 2$. It is also observed that when long jumps are frequent the search is much faster, although the gain in
search efficiency due to the communication mechanisms is lower close to the ballistic limit (i.e., $\mu\rightarrow 0$).
This will be explained later in Section \ref{comparation}.

Similarly to the Brownian case, a particularly simple limit in Eq.~(\ref{diffsetlevy})
appears when 
$\tau_{+}>>\tau_{-}$ for $x>L/2$ (and the contrary on the 
other half of the system). The search time is 
\begin{equation}
T(x=L,\sigma)\propto\frac{1}{\tau_{+}}, \nonumber \\
\end{equation}
where the proportionality constant is a combination of the coefficients $A_i$ that depends on the size of the system.

\begin{figure}
\begin{center} 
\includegraphics[width=0.45\textwidth]{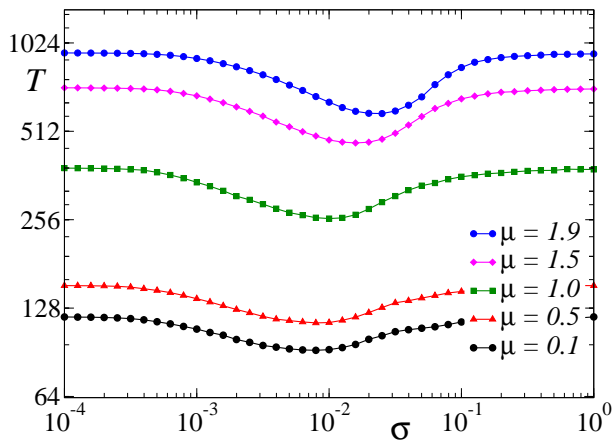}
\caption{(Color online). Mean first arrival time for L\'evy flights with different exponent $\mu$ in the $2-D$ model.
$B_{g}=1$, $B_{c}=1$, $\tau_{0}=50$. Lines are interpolations.}
\label{comp}
\end{center}
\end{figure}

{\it Results in two dimensions.}
We now present some results in $2-D$
using Monte Carlo simulations, as was done in the case of Brownian particles in \cite{Martinez-Garcia2013b}. 
The individuals are moving on a discrete regular square lattice $(L_{x}=L_{y}=1)$ of mesh $h=0.01$, where the 
targets are randomly distributed. Similarly to  the $1-D$ case, the searchers stop if they find a target
during a displacement of length $l$.
This naturally introduces a cutoff in the length of the jumps, which becomes more important as target density increases \cite{libroforaging}. 
However, as we will focus on a situation where target density is low, we introduce an exponential cutoff of the order of the system size in the jump length
probability distribution to ensure that very long jumps 
without physical meaning (they imply very high velocities)
do not occur
\begin{equation}\label{cutoffdist}
\varphi_{\mu}(l)=C\frac{\exp(-l/L)\mu b^{1/\mu}}{\tilde{l}\left(1+\frac{l}{\tilde{l}} b^{1/\mu} \right)^{\mu+1}},
\end{equation}
where $C=\int_{0}^{\infty}\varphi_{\mu}(l)dl$ is the normalization constant, and $\tilde{l}=h$. 
We did not need to introduce such a cutoff in the study of the $1-D$ model because the boundaries of the system introduced a natural 
truncation in the jump length distribution, and jumps longer than the system size never occurred.
Generally, the search is faster when long displacements occur more frequently.
Figure \ref{comp} shows search time for different values of $\mu$.
Again, the effect of the communication mechanism is more important when
we approach the Brownian limit ($\mu\rightarrow 2$), as will be 
explained next. 

\section{Influence of the searching strategy: L\'evy vs Brownian}\label{comparation}

As a general result of the model, searching is faster when individuals have intermediate 
amounts of information, regardless of the kind of movement strategy
followed by the population (Brownian or L\'evy). However,
communication has a larger impact on Brownian motion, i.e.,
the depth of the well at $\sigma_{opt}$ is larger (Figures \ref{timesana} and \ref{comp}).

\begin{figure}
\begin{center} 
\includegraphics[width=0.45\textwidth]{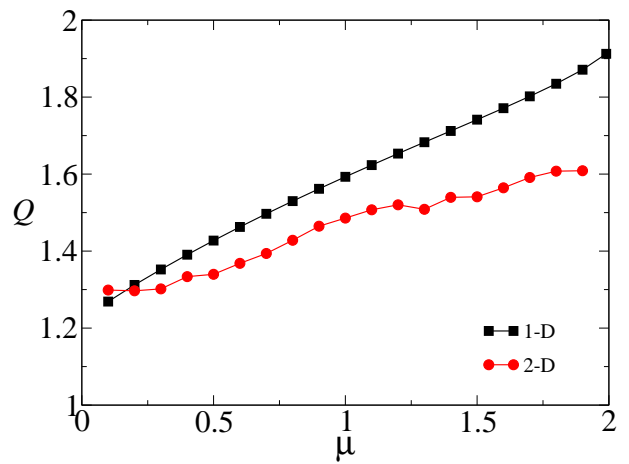}
\caption{(Color online). Improvement of the searching process because of the communication mechanism. Circles correspond 
to the $2-D$ model and squares to $1-D$. Lines are interpolations.}
\label{mejora}
\end{center}
\end{figure}

A measure of the improvement in search performance at the optimal communication range is given
by the ratio between the search time without communication and that at 
the optimal communication range, $Q=T_{\sigma\rightarrow0}/T_{\sigma_{opt}}$. 
This quantity is plotted in Figure \ref{mejora} for different L\'evy exponents.
As previously mentioned, Brownian searchers that are not able to perform long displacements benefit more
from communication than L\'evy searchers. This is because introducing an additional source of information
increases the directionality of the random motion and prevents the searcher from revisiting the same place many times, which
is the key problem with Brownian search strategies \cite{libroforaging}. A Brownian walker
has no directionality in the movement, so provided with sources of information (communication together with the local 
quality of the landscape) it can search much more efficiently. This effect is less important
for L\'evy searchers due to the presence of long, straight-line moves that, by themselves, decrease the number
of times that a particular area is revisited.
In summary,  the
communication mechanism is less important in L\'evy strategies, so that its effect is less noticeable
as it is shown in Figure \ref{mejora} both in $1-D$ and $2-D$.

However, the value of the optimal interaction range
changes with the kind of motion. This is shown in the $2-D$ model by the dependence of the mean
search time on the communication range for different L\'evy exponents (Figure \ref{comp}).
The value of $\sigma_{opt}$ increases with the L\'evy exponent, 
so Brownian searchers $(\mu\rightarrow2)$ need
to spread the information farther (a larger value of $\sigma_{opt}$)
than L\'evy $(\mu=1)$ walkers to obtain the maximum benefit.
L\'evy trajectories show clusters of short displacements with frequent turns occasionally broken up by long linear displacements,
which account for most of the target encounters. However, because these steps are often much longer than the average distance between targets
they are not positively influenced by communication, so any benefit a L\'evy strategy gains from communication occurs during the series
of short displacements. The time that an individual spends doing short movements is limited by the interarrival time of the large steps, so 
unless an individual is already relatively close to a target, it will not have time to reach a target before the next big step comes and
moves it far away from that original target. Therefore the
optimal communication range decreases with the L\'evy exponent, $\mu$, as longer displacements become more frequent at lower $\mu$ values. 
 
In addition, the value of $\sigma_{opt}$ depends on both the number of targets and their spatial distribution,
as was shown at the end of Section \ref{brownsec} for a simple $1-D$ situation where $\sigma_{opt}\sim L$.

\section{Summary and conclusions}\label{summmary}

In this paper we compared
Brownian and L\'evy search strategies using a population of individuals 
that exchange information about the location of spatially distributed targets. 
Using a simple $1-D$ model we have provided analytical results on both cases, concluding that
frequent long jumps ($\mu\rightarrow 0$, ballistic limit) minimize the searching times.

 However the effect of a communication mechanism is more pronounced in the limit of short 
jumps i.e., Brownian motion. This means that a population of individuals employing Brownian motion
gains proportionally more benefit from communicating and sharing information than does a population of
L\'evy walkers, where long jumps are more or less frequent depending on the value of the L\'evy exponent $\mu$.
When messages are exchanged in a range that minimizes search
duration, communication is the driving force in the Brownian limit, but 
occasional long jumps are still responsible for most of the encounters with targets in the case of long-tailed
step-length distributions.

The  main result of this work is rather general: 
independently of the kind of communication performed by the population, and of the spatial
distribution of the targets, a population of individuals with the ability to communicate will find the 
targets in a shorter time if the information is spread at intermediate ranges. Both an excess and a lack of information increase the search time.
However, the communication mechanism does not have the same quantitative effect on the different moving strategies (i.e., ballistic, L\'evy or Brownian).
Uninformed Brownian individuals perform a random movement revisiting the same position many times, so having an external source of information
introduces directionality on the movement, decreasing the number of times that a point in the space is visited. In the case of L\'evy and ballistic 
strategies $(\mu\rightarrow0)$, communication is less noticeable because individuals are able to do long jumps.
This is already a source of directionality that prevents individuals from revisiting 
the same points in space many times, and thus weakening the effect of the directionality introduced by communication.

\section{Acknowledgments}
R.M-G. is supported by the JAEPredoc program of CSIC.
R.M-G. and C.L acknowledge support from MINECO
(Spain) and FEDER (EU) through Grants No. FIS2012-30634 (Intense-COSYP)
and CTM2012-39025-C02-01 (ESCOLA). J.M.C. is supported by US National Science Foundation grant ABI 1062411. 
We thank Federico Vazquez for fruitful discussions.
 
\appendix

\section{Voronoi diagrams of the model}

\begin{figure}
\begin{center} 
\includegraphics[width=.45\textwidth]{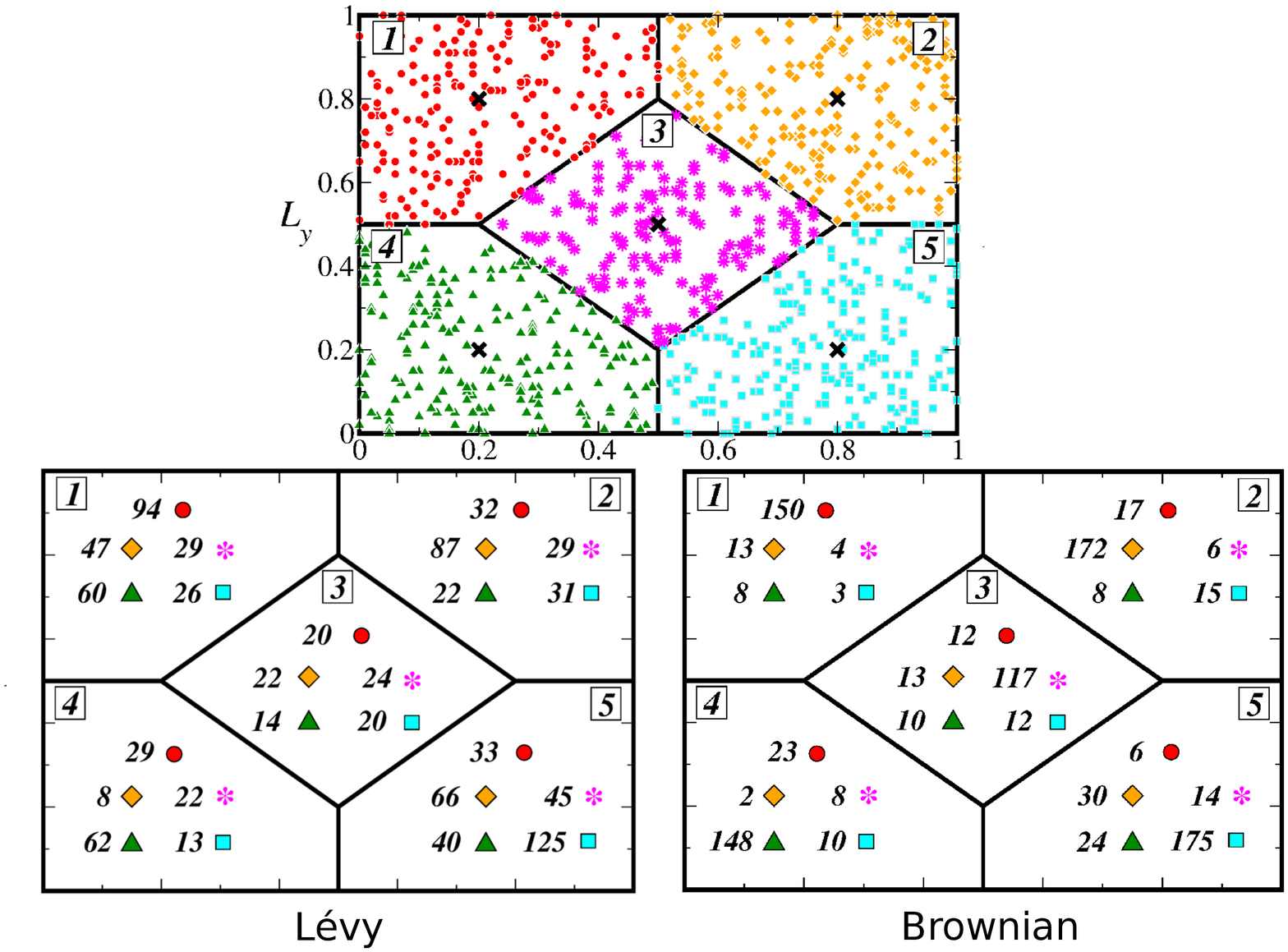}
\caption{(Color online). (Top) Initial random distribution of individuals, the symbol refers to the Voronoi cell at which every individual 
belongs initially.  (Bottom left) Number of individuals coming from each cell at each target at the end of the search using L\'evy flights.
(Right) Number of individuals coming from each cell
 at each target at the end of the search using Brownian motion. Parameters: $\sigma=0.01$ (optimal communication
range), $B_{g}=1$, $B_{c}=1$, $\tau_{0}=50$. The black crosses represent the location of the $5$ targets.}
\label{voronoi}
\end{center}
\end{figure}

The behavior of the model, resulting in optimal searches at intermediate communication ranges,
can be explained in terms of Voronoi diagrams \cite{voronoi}. Consider every target as a seed that has 
associated a Voronoi cell formed by those points whose distance to that seed is less than or equal to
its distance to any other one (See Figure \ref{voronoi} (top) for a distribution
of the space in $5$ Voronoi cells for an initial distribution of particles with five targets (crosses)).
The searching time will be minimized when the information coming from the individuals located on one
target covers the full associated Voronoi cell, but only that cell.
In this situation, the searchers within that cell will receive information coming only from that target
 and move towards it. $\sigma_{opt}$ is the communication range that maximizes the gradient 
 (approximately the smallest value of $\sigma$ that makes
 the calling function not vanishing) of 
 the calling function at the frontiers of the Voronoi cells.
 Increasing the communication range provides individuals with information 
 coming from different targets, and makes them get overwhelmed in the limit $\sigma\rightarrow\infty$.
This Voronoi construction may also help to explain
the improvement of the searching strategies because of sharing
information.
The difference between Brownian and L\'evy strategies can be seen 
 in Figure \ref{voronoi} (Bottom).
They show the origin of the individuals that are at each target at the end of a L\'evy (Left) and a 
Brownian search (Right) (i.e., in which Voronoi cell they were at the beginning).
In the case of Brownian individuals most of the particles at every target were initially in its Voronoi cell. For
L\'evy flights the long displacements mix the population in the stationary state
(i.e., individuals at a target come from different cells). The 
communication mechanism is less important in L\'evy strategies, so that its effect is less noticeable and the encounters
of individuals with targets are caused mainly by the long displacements.

\section{Deterministic approach to the searching times.}\label{app2}

It is possible to improve the results given by the deterministic approach if the region
close to the target, i.e. the boundary of the system, is neglected
in the average given by Eq.~(\ref{average}). At that point, one of the rates is much higher than the other 
and thus would contribute to the drift velocity making its value much higher, 
mainly in the limit $\sigma\rightarrow 0$.
To this aim one can include a parameter $\epsilon$, so that the integration limits
in Eq.~(\ref{average}) are $L/2$ and $L-\epsilon$.

To estimate the value of $\epsilon$ it is useful to plot
$\tau_{+}(x)-\tau_{-}(x)$ versus $x$ (not shown). The difference between rates, although 
depending on $\sigma$, starts increasing quickly when $x\geq L-2$, so one can estimate $\epsilon=2$. 
The inset of Figure \ref{timesana} shows the exit time as a function of the communication range computed with this approach (dashed line).
Its optimal value is in good agreement with the result obtained using the definition
of the search time (thick line), with $\sigma_{opt}\approx12.5$ for both approaches.
However the temporal scale of the problem (the absolute values of the times), although higher
than with $\epsilon=0$,  is still lower in this calculation.


\begin{thebibliography}{37}
\expandafter\ifx\csname natexlab\endcsname\relax\def\natexlab#1{#1}\fi
\expandafter\ifx\csname bibnamefont\endcsname\relax
  \def\bibnamefont#1{#1}\fi
\expandafter\ifx\csname bibfnamefont\endcsname\relax
  \def\bibfnamefont#1{#1}\fi
\expandafter\ifx\csname citenamefont\endcsname\relax
  \def\citenamefont#1{#1}\fi
\expandafter\ifx\csname url\endcsname\relax
  \def\url#1{\texttt{#1}}\fi
\expandafter\ifx\csname urlprefix\endcsname\relax\def\urlprefix{URL }\fi
\providecommand{\bibinfo}[2]{#2}
\providecommand{\eprint}[2][]{\url{#2}}

\bibitem[{\citenamefont{B\'{e}nichou et~al.}(2011)\citenamefont{B\'{e}nichou,
  Loverdo, Moreau, and Voituriez}}]{rmp_benichou}
\bibinfo{author}{\bibfnamefont{O.}~\bibnamefont{B\'{e}nichou}},
  \bibinfo{author}{\bibfnamefont{C.}~\bibnamefont{Loverdo}},
  \bibinfo{author}{\bibfnamefont{M.}~\bibnamefont{Moreau}}, \bibnamefont{and}
  \bibinfo{author}{\bibfnamefont{R.}~\bibnamefont{Voituriez}},
  \bibinfo{journal}{Reviews of Modern Physics} \textbf{\bibinfo{volume}{83}},
  \bibinfo{pages}{81} (\bibinfo{year}{2011}).

\bibitem[{\citenamefont{M\'{e}ndez et~al.}(2014)\citenamefont{M\'{e}ndez,
  Campos, and Bartumeus}}]{mendez2014random}
\bibinfo{author}{\bibfnamefont{V.}~\bibnamefont{M\'{e}ndez}},
  \bibinfo{author}{\bibfnamefont{D.}~\bibnamefont{Campos}}, \bibnamefont{and}
  \bibinfo{author}{\bibfnamefont{F.}~\bibnamefont{Bartumeus}}, in
  \emph{\bibinfo{booktitle}{Stochastic Foundations in Movement Ecology}}
  (\bibinfo{publisher}{Springer Berlin Heidelberg}, \bibinfo{year}{2014}), pp.
  \bibinfo{pages}{177--205}.

\bibitem[{\citenamefont{Pirolli and Card}(1999)}]{pirolli}
\bibinfo{author}{\bibfnamefont{P.}~\bibnamefont{Pirolli}} \bibnamefont{and}
  \bibinfo{author}{\bibfnamefont{S.}~\bibnamefont{Card}},
  \bibinfo{journal}{Psychological Review} \textbf{\bibinfo{volume}{106}},
  \bibinfo{pages}{643} (\bibinfo{year}{1999}).

\bibitem[{\citenamefont{Viswanathan et~al.}(2011)\citenamefont{Viswanathan,
  da~Luz, Raposo, and Stanley}}]{libroforaging}
\bibinfo{author}{\bibfnamefont{G.~M}~\bibnamefont{Viswanathan}},
  \bibinfo{author}{\bibfnamefont{M.~G.~E.} \bibnamefont{da~Luz}},
  \bibinfo{author}{\bibfnamefont{E.~P.} \bibnamefont{Raposo}},
  \bibnamefont{and} \bibinfo{author}{\bibfnamefont{H.~E.}
  \bibnamefont{Stanley}}, \emph{\bibinfo{title}{{The physics of foraging: an
  introduction to random searches and biological encounters}}}
  (\bibinfo{publisher}{Cambridge University Press}, \bibinfo{year}{2011}),
  \bibinfo{edition}{1st} ed.

\bibitem[{\citenamefont{Vergassola et~al.}(2007)\citenamefont{Vergassola,
  Villermaux, and Shraiman}}]{nature-vergassola}
\bibinfo{author}{\bibfnamefont{M.}~\bibnamefont{Vergassola}},
  \bibinfo{author}{\bibfnamefont{E.}~\bibnamefont{Villermaux}},
  \bibnamefont{and} \bibinfo{author}{\bibfnamefont{B.~I.}
  \bibnamefont{Shraiman}}, \bibinfo{journal}{Nature}
  \textbf{\bibinfo{volume}{445}}, \bibinfo{pages}{406} (\bibinfo{year}{2007}).

\bibitem[{\citenamefont{B\'{e}nichou et~al.}(2006)\citenamefont{B\'{e}nichou,
  Loverdo, Moreau, and Voituriez}}]{benPRE2006}
\bibinfo{author}{\bibfnamefont{O.}~\bibnamefont{B\'{e}nichou}},
  \bibinfo{author}{\bibfnamefont{C.}~\bibnamefont{Loverdo}},
  \bibinfo{author}{\bibfnamefont{M.}~\bibnamefont{Moreau}}, \bibnamefont{and}
  \bibinfo{author}{\bibfnamefont{R.}~\bibnamefont{Voituriez}},
  \bibinfo{journal}{Physical Review E} \textbf{\bibinfo{volume}{74}},
  \bibinfo{pages}{020102} (\bibinfo{year}{2006}).

\bibitem[{\citenamefont{Taylor and Halford}(1989)}]{bio-taylor}
\bibinfo{author}{\bibfnamefont{J.~D.} \bibnamefont{Taylor}} \bibnamefont{and}
  \bibinfo{author}{\bibfnamefont{S.~E.} \bibnamefont{Halford}},
  \bibinfo{journal}{Biochemistry} \textbf{\bibinfo{volume}{28}},
  \bibinfo{pages}{6198} (\bibinfo{year}{1989}).

\bibitem[{\citenamefont{Campos et~al.}(2013)\citenamefont{Campos, Bartumeus,
  and M\'endez}}]{Campos2013}
\bibinfo{author}{\bibfnamefont{D.}~\bibnamefont{Campos}},
  \bibinfo{author}{\bibfnamefont{F.}~\bibnamefont{Bartumeus}},
  \bibnamefont{and} \bibinfo{author}{\bibfnamefont{V.}
  \bibnamefont{M\'endez}}, \bibinfo{journal}{Phys. Rev. E}
  \textbf{\bibinfo{volume}{88}}, \bibinfo{pages}{022101}
  (\bibinfo{year}{2013}).

\bibitem[{\citenamefont{Viswanathan et~al.}(1999)\citenamefont{Viswanathan,
  Buldyrev, Havlin, da~Luz, Raposo, and Stanley}}]{visna1999}
\bibinfo{author}{\bibfnamefont{G.~M.} \bibnamefont{Viswanathan}},
  \bibinfo{author}{\bibfnamefont{S.~V.} \bibnamefont{Buldyrev}},
  \bibinfo{author}{\bibfnamefont{S.}~\bibnamefont{Havlin}},
  \bibinfo{author}{\bibfnamefont{M.~G.} \bibnamefont{da~Luz}},
  \bibinfo{author}{\bibfnamefont{E.~P.} \bibnamefont{Raposo}},
  \bibnamefont{and} \bibinfo{author}{\bibfnamefont{H.~E.}
  \bibnamefont{Stanley}}, \bibinfo{journal}{Nature}
  \textbf{\bibinfo{volume}{401}}, \bibinfo{pages}{911} (\bibinfo{year}{1999}).

\bibitem[{\citenamefont{Shlesinger}(2006)}]{shlesinger2006}
\bibinfo{author}{\bibfnamefont{M.~F.} \bibnamefont{Shlesinger}},
  \bibinfo{journal}{Nature} \textbf{\bibinfo{volume}{443}},
  \bibinfo{pages}{281} (\bibinfo{year}{2006}).

\bibitem[{\citenamefont{Edwards et~al.}(2007)\citenamefont{Edwards, Phillips,
  Watkins, Freeman, Murphy, Afanasyev, Buldyrev, da~Luz, Raposo, Stanley
  et~al.}}]{edwards2007}
\bibinfo{author}{\bibfnamefont{A.~M.} \bibnamefont{Edwards}},
  \bibinfo{author}{\bibfnamefont{R.~A.} \bibnamefont{Phillips}},
  \bibinfo{author}{\bibfnamefont{N.~W.} \bibnamefont{Watkins}},
  \bibinfo{author}{\bibfnamefont{M.~P.} \bibnamefont{Freeman}},
  \bibinfo{author}{\bibfnamefont{E.~J.} \bibnamefont{Murphy}},
  \bibinfo{author}{\bibfnamefont{V.}~\bibnamefont{Afanasyev}},
  \bibinfo{author}{\bibfnamefont{S.~V.} \bibnamefont{Buldyrev}},
  \bibinfo{author}{\bibfnamefont{M.~G.~E.} \bibnamefont{da~Luz}},
  \bibinfo{author}{\bibfnamefont{E.~P.} \bibnamefont{Raposo}},
  \bibinfo{author}{\bibfnamefont{H.~E.} \bibnamefont{Stanley}},
  \bibnamefont{et~al.}, \bibinfo{journal}{Nature}
  \textbf{\bibinfo{volume}{449}}, \bibinfo{pages}{1044} (\bibinfo{year}{2007}).

\bibitem[{\citenamefont{Torney et~al.}(2011)\citenamefont{Torney, Berdahl, and
  Couzin}}]{Torney2011}
\bibinfo{author}{\bibfnamefont{C.~J.} \bibnamefont{Torney}},
  \bibinfo{author}{\bibfnamefont{A.}~\bibnamefont{Berdahl}}, \bibnamefont{and}
  \bibinfo{author}{\bibfnamefont{I.~D.} \bibnamefont{Couzin}},
  \bibinfo{journal}{PLoS computational biology} \textbf{\bibinfo{volume}{7}},
  \bibinfo{pages}{e1002194} (\bibinfo{year}{2011}).

\bibitem[{\citenamefont{Hein and McKinley}(2012)}]{hein}
\bibinfo{author}{\bibfnamefont{A.~M.} \bibnamefont{Hein}} \bibnamefont{and}
  \bibinfo{author}{\bibfnamefont{S.~A.} \bibnamefont{McKinley}},
  \bibinfo{journal}{Proceedings of the National Academy of Sciences of the
  United States of America} \textbf{\bibinfo{volume}{109}},
  \bibinfo{pages}{12070} (\bibinfo{year}{2012}).

\bibitem[{\citenamefont{Viswanathan et~al.}(2008)\citenamefont{Viswanathan,
  Raposo, and da~Luz}}]{Viswanathan2008}
\bibinfo{author}{\bibfnamefont{G.}~\bibnamefont{Viswanathan}},
  \bibinfo{author}{\bibfnamefont{E.}~\bibnamefont{Raposo}}, \bibnamefont{and}
  \bibinfo{author}{\bibfnamefont{M.}~\bibnamefont{da~Luz}},
  \bibinfo{journal}{Physics of Life Reviews} \textbf{\bibinfo{volume}{5}},
  \bibinfo{pages}{133} (\bibinfo{year}{2008}).

\bibitem[{\citenamefont{Mej\'{\i}a-Monasterio
  et~al.}(2011)\citenamefont{Mej\'{\i}a-Monasterio, Oshanin, and
  Schehr}}]{mejiamonasterio}
\bibinfo{author}{\bibfnamefont{C.}~\bibnamefont{Mej\'{\i}a-Monasterio}},
  \bibinfo{author}{\bibfnamefont{G.}~\bibnamefont{Oshanin}}, \bibnamefont{and}
  \bibinfo{author}{\bibfnamefont{G.}~\bibnamefont{Schehr}},
  \bibinfo{journal}{Journal of Statistical Mechanics: Theory and Experiment}
  \textbf{\bibinfo{volume}{2011}}, \bibinfo{pages}{P06022}
  (\bibinfo{year}{2011}).

\bibitem[{\citenamefont{Bartumeus et~al.}(2003)\citenamefont{Bartumeus, Peters,
  Pueyo, Marras\'{e}, and Catalan}}]{bartomeus}
\bibinfo{author}{\bibfnamefont{F.}~\bibnamefont{Bartumeus}},
  \bibinfo{author}{\bibfnamefont{F.}~\bibnamefont{Peters}},
  \bibinfo{author}{\bibfnamefont{S.}~\bibnamefont{Pueyo}},
  \bibinfo{author}{\bibfnamefont{C.}~\bibnamefont{Marras\'{e}}},
  \bibnamefont{and} \bibinfo{author}{\bibfnamefont{J.}~\bibnamefont{Catalan}},
  \bibinfo{journal}{Proceedings of the National Academy of Sciences of the
  United States of America} \textbf{\bibinfo{volume}{100}},
  \bibinfo{pages}{12771} (\bibinfo{year}{2003}).

\bibitem[{\citenamefont{Bartumeus et~al.}(2005)\citenamefont{Bartumeus, da~Luz,
  Viswanathan, and Catalan}}]{bartumeus2005animal}
\bibinfo{author}{\bibfnamefont{F.}~\bibnamefont{Bartumeus}},
  \bibinfo{author}{\bibfnamefont{M.~G.~E.} \bibnamefont{da~Luz}},
  \bibinfo{author}{\bibfnamefont{G.}~\bibnamefont{Viswanathan}},
  \bibnamefont{and} \bibinfo{author}{\bibfnamefont{J.}~\bibnamefont{Catalan}},
  \bibinfo{journal}{Ecology} \textbf{\bibinfo{volume}{86}},
  \bibinfo{pages}{3078} (\bibinfo{year}{2005}).

\bibitem[{\citenamefont{Bartumeus et~al.}(2002)\citenamefont{Bartumeus,
  Catalan, Fulco, Lyra, and Viswanathan}}]{BartumeusPRL}
\bibinfo{author}{\bibfnamefont{F.}~\bibnamefont{Bartumeus}},
  \bibinfo{author}{\bibfnamefont{J.}~\bibnamefont{Catalan}},
  \bibinfo{author}{\bibfnamefont{U.~L}~\bibnamefont{Fulco}},
  \bibinfo{author}{\bibfnamefont{M.~L}~\bibnamefont{Lyra}}, \bibnamefont{and}
  \bibinfo{author}{\bibfnamefont{G.~M}~\bibnamefont{Viswanathan}},
  \bibinfo{journal}{Physical Review Letters} \textbf{\bibinfo{volume}{88}},
  \bibinfo{pages}{097901} (\bibinfo{year}{2002}).

\bibitem[{\citenamefont{Hoare et~al.}(2004)\citenamefont{Hoare, Couzin, Godin,
  and Krause}}]{hoare}
\bibinfo{author}{\bibfnamefont{D.}~\bibnamefont{Hoare}},
  \bibinfo{author}{\bibfnamefont{I.}~\bibnamefont{Couzin}},
  \bibinfo{author}{\bibfnamefont{J.-G.} \bibnamefont{Godin}}, \bibnamefont{and}
  \bibinfo{author}{\bibfnamefont{J.}~\bibnamefont{Krause}},
  \bibinfo{journal}{Animal Behaviour} \textbf{\bibinfo{volume}{67}},
  \bibinfo{pages}{155} (\bibinfo{year}{2004}).

\bibitem[{\citenamefont{Torney et~al.}(2009)\citenamefont{Torney, Neufeld, and
  Couzin}}]{torneypnas}
\bibinfo{author}{\bibfnamefont{C.}~\bibnamefont{Torney}},
  \bibinfo{author}{\bibfnamefont{Z.}~\bibnamefont{Neufeld}}, \bibnamefont{and}
  \bibinfo{author}{\bibfnamefont{I.~D.} \bibnamefont{Couzin}},
  \bibinfo{journal}{Proceedings of the National Academy of Sciences of the
  United States of America} \textbf{\bibinfo{volume}{106}},
  \bibinfo{pages}{22055} (\bibinfo{year}{2009}).

\bibitem[{\citenamefont{Viswanathan et~al.}(2000)\citenamefont{Viswanathan,
  Afanasyev, Buldyrev, Havlin, da~Luz, Raposo, and Stanley}}]{viswaphysica}
\bibinfo{author}{\bibfnamefont{G.~M}~\bibnamefont{Viswanathan}},
  \bibinfo{author}{\bibfnamefont{V.}~\bibnamefont{Afanasyev}},
  \bibinfo{author}{\bibfnamefont{S.~V.} \bibnamefont{Buldyrev}},
  \bibinfo{author}{\bibfnamefont{S.}~\bibnamefont{Havlin}},
  \bibinfo{author}{\bibfnamefont{M.}~\bibnamefont{da~Luz}},
  \bibinfo{author}{\bibfnamefont{E.}~\bibnamefont{Raposo}}, \bibnamefont{and}
  \bibinfo{author}{\bibfnamefont{H.}~\bibnamefont{Stanley}},
  \bibinfo{journal}{Physica A: Statistical Mechanics and its Applications}
  \textbf{\bibinfo{volume}{282}}, \bibinfo{pages}{1} (\bibinfo{year}{2000}).

\bibitem[{\citenamefont{Edwards}(2011)}]{edwards2011overturning}
\bibinfo{author}{\bibfnamefont{A.~M.} \bibnamefont{Edwards}},
  \bibinfo{journal}{Ecology} \textbf{\bibinfo{volume}{92}},
  \bibinfo{pages}{1247} (\bibinfo{year}{2011}).

\bibitem[{\citenamefont{Liu and Passino}(2002)}]{LiuPassino}
\bibinfo{author}{\bibfnamefont{Y.}~\bibnamefont{Liu}} \bibnamefont{and}
  \bibinfo{author}{\bibfnamefont{K.}~\bibnamefont{Passino}},
  \bibinfo{journal}{Journal of Optimization Theory and Applications}
  \textbf{\bibinfo{volume}{115}}, \bibinfo{pages}{603} (\bibinfo{year}{2002}).

\bibitem[{\citenamefont{Zuberb\"{u}hler
  et~al.}(1997)\citenamefont{Zuberb\"{u}hler, No\"{e}, and
  Seyfarth}}]{dianamonkey}
\bibinfo{author}{\bibfnamefont{K.}~\bibnamefont{Zuberb\"{u}hler}},
  \bibinfo{author}{\bibfnamefont{R.}~\bibnamefont{No\"{e}}}, \bibnamefont{and}
  \bibinfo{author}{\bibfnamefont{R.}~\bibnamefont{Seyfarth}},
  \bibinfo{journal}{Animal Behaviour} \textbf{\bibinfo{volume}{53}}, \bibinfo{pages}{589--604}
  (\bibinfo{year}{1997}).

\bibitem[{\citenamefont{McComb et~al.}(2003)\citenamefont{McComb, Reby, Baker,
  Moss, and Sayialel}}]{mccomb}
\bibinfo{author}{\bibfnamefont{K.}~\bibnamefont{McComb}},
  \bibinfo{author}{\bibfnamefont{D.}~\bibnamefont{Reby}},
  \bibinfo{author}{\bibfnamefont{L.}~\bibnamefont{Baker}},
  \bibinfo{author}{\bibfnamefont{C.}~\bibnamefont{Moss}}, \bibnamefont{and}
  \bibinfo{author}{\bibfnamefont{S.}~\bibnamefont{Sayialel}},
  \bibinfo{journal}{Animal Behaviour} \textbf{\bibinfo{volume}{65}},
  \bibinfo{pages}{317} (\bibinfo{year}{2003}).

\bibitem[{\citenamefont{Mishra et~al.}(2012)\citenamefont{Mishra, Tunstr\o{}m,
  Couzin, and Huepe}}]{PhysRevE.86.011901}
\bibinfo{author}{\bibfnamefont{S.}~\bibnamefont{Mishra}},
  \bibinfo{author}{\bibfnamefont{K.}~\bibnamefont{Tunstr\o{}m}},
  \bibinfo{author}{\bibfnamefont{I.~D.} \bibnamefont{Couzin}}, \bibnamefont{and}
  \bibinfo{author}{\bibfnamefont{C.}~\bibnamefont{Huepe}},
  \bibinfo{journal}{Phys. Rev. E} \textbf{\bibinfo{volume}{86}},
  \bibinfo{pages}{011901} (\bibinfo{year}{2012}).

\bibitem[{\citenamefont{Kolpas et~al.}(2013)\citenamefont{Kolpas, Busch, Li,
  Couzin, Petzold, and Moehlis}}]{Kolpas2013}
\bibinfo{author}{\bibfnamefont{A.}~\bibnamefont{Kolpas}},
  \bibinfo{author}{\bibfnamefont{M.}~\bibnamefont{Busch}},
  \bibinfo{author}{\bibfnamefont{H.}~\bibnamefont{Li}},
  \bibinfo{author}{\bibfnamefont{I.~D.} \bibnamefont{Couzin}},
  \bibinfo{author}{\bibfnamefont{L.}~\bibnamefont{Petzold}}, \bibnamefont{and}
  \bibinfo{author}{\bibfnamefont{J.}~\bibnamefont{Moehlis}},
  \bibinfo{journal}{PloS one} \textbf{\bibinfo{volume}{8}},
  \bibinfo{pages}{e58525} (\bibinfo{year}{2013}).

\bibitem[{\citenamefont{Couzin et~al.}(2002)\citenamefont{Couzin, Krause,
  James, Ruxton, and Franks}}]{Couzin2002}
\bibinfo{author}{\bibfnamefont{I.~D.} \bibnamefont{Couzin}},
  \bibinfo{author}{\bibfnamefont{J.}~\bibnamefont{Krause}},
  \bibinfo{author}{\bibfnamefont{R.}~\bibnamefont{James}},
  \bibinfo{author}{\bibfnamefont{G.~D.} \bibnamefont{Ruxton}},
  \bibnamefont{and} \bibinfo{author}{\bibfnamefont{N.~R.}
  \bibnamefont{Franks}}, \bibinfo{journal}{Journal of Theoretical Biology}
  \textbf{\bibinfo{volume}{218}}, \bibinfo{pages}{1} (\bibinfo{year}{2002}).

\bibitem[{\citenamefont{Mart\'{\i}nez-Garc\'{\i}a
  et~al.}(2013)\citenamefont{Mart\'{\i}nez-Garc\'{\i}a, Calabrese, Mueller,
  Olson, and L\'{o}pez}}]{Martinez-Garcia2013b}
\bibinfo{author}{\bibfnamefont{R.}~\bibnamefont{Mart\'{\i}nez-Garc\'{\i}a}},
  \bibinfo{author}{\bibfnamefont{J.~M.} \bibnamefont{Calabrese}},
  \bibinfo{author}{\bibfnamefont{T.}~\bibnamefont{Mueller}},
  \bibinfo{author}{\bibfnamefont{K.~A.} \bibnamefont{Olson}}, \bibnamefont{and}
  \bibinfo{author}{\bibfnamefont{C.}~\bibnamefont{L\'{o}pez}},
  \bibinfo{journal}{Physical Review Letters} \textbf{\bibinfo{volume}{110}},
  \bibinfo{pages}{248106} (\bibinfo{year}{2013}).

\bibitem[{\citenamefont{Smouse et~al.}(2010)\citenamefont{Smouse, Focardi,
  Moorcroft, Kie, Forester, and Morales}}]{smouse}
\bibinfo{author}{\bibfnamefont{P.~E.} \bibnamefont{Smouse}},
  \bibinfo{author}{\bibfnamefont{S.}~\bibnamefont{Focardi}},
  \bibinfo{author}{\bibfnamefont{P.~R.} \bibnamefont{Moorcroft}},
  \bibinfo{author}{\bibfnamefont{J.~G.} \bibnamefont{Kie}},
  \bibinfo{author}{\bibfnamefont{J.~D.} \bibnamefont{Forester}},
  \bibnamefont{and} \bibinfo{author}{\bibfnamefont{J.~M.}
  \bibnamefont{Morales}}, \bibinfo{journal}{Philosophical transactions of the
  Royal Society of London. Series B, Biological sciences}
  \textbf{\bibinfo{volume}{365}}, \bibinfo{pages}{2201} (\bibinfo{year}{2010}).

\bibitem[{\citenamefont{Preisler et~al.}(2004)\citenamefont{Preisler, Ager,
  Johnson, and Kie}}]{Preisler}
\bibinfo{author}{\bibfnamefont{H.~K.} \bibnamefont{Preisler}},
  \bibinfo{author}{\bibfnamefont{A.~A.} \bibnamefont{Ager}},
  \bibinfo{author}{\bibfnamefont{B.~K.} \bibnamefont{Johnson}},
  \bibnamefont{and} \bibinfo{author}{\bibfnamefont{J.~G.} \bibnamefont{Kie}},
  \bibinfo{journal}{Environmetrics} \textbf{\bibinfo{volume}{15}},
  \bibinfo{pages}{643} (\bibinfo{year}{2004}).

\bibitem[{\citenamefont{Turchin}(1998)}]{Turchin}
\bibinfo{author}{\bibfnamefont{P.}~\bibnamefont{Turchin}},
  \emph{\bibinfo{title}{{Quantitative Analysis of Movement: measuring and
  modeling population redistribution in animals and plants.}}}
  (\bibinfo{publisher}{Sinauer Associates, Incorporated}, \bibinfo{year}{1998}).

\bibitem[{\citenamefont{Dean}(1996)}]{Dean}
\bibinfo{author}{\bibfnamefont{D.~S.} \bibnamefont{Dean}},
  \bibinfo{journal}{Journal of Physics A: Mathematical and General}
  \textbf{\bibinfo{volume}{29}}, \bibinfo{pages}{L613} (\bibinfo{year}{1996}).

\bibitem[{\citenamefont{Redner}(2001)}]{redner}
\bibinfo{author}{\bibfnamefont{S.}~\bibnamefont{Redner}},
  \emph{\bibinfo{title}{{A guide to first passage processes}}}
  (\bibinfo{publisher}{Cambridge University Press}, \bibinfo{year}{2001}).

\bibitem[{\citenamefont{Humphries et~al.}(2010)\citenamefont{Humphries,
  Queiroz, Dyer, Pade, Musyl, Schaefer, Fuller, Brunnschweiler, Doyle, Houghton
  et~al.}}]{nature_hum}
\bibinfo{author}{\bibfnamefont{N.~E.} \bibnamefont{Humphries}},
  \bibinfo{author}{\bibfnamefont{N.}~\bibnamefont{Queiroz}},
  \bibinfo{author}{\bibfnamefont{J.~R.~M.} \bibnamefont{Dyer}},
  \bibinfo{author}{\bibfnamefont{N.~G.} \bibnamefont{Pade}},
  \bibinfo{author}{\bibfnamefont{M.~K.} \bibnamefont{Musyl}},
  \bibinfo{author}{\bibfnamefont{K.~M.} \bibnamefont{Schaefer}},
  \bibinfo{author}{\bibfnamefont{D.~W.} \bibnamefont{Fuller}},
  \bibinfo{author}{\bibfnamefont{J.~M.} \bibnamefont{Brunnschweiler}},
  \bibinfo{author}{\bibfnamefont{T.~K.} \bibnamefont{Doyle}},
  \bibinfo{author}{\bibfnamefont{J.~D.~R.} \bibnamefont{Houghton}},
  \bibnamefont{et~al.}, \bibinfo{journal}{Nature}
  \textbf{\bibinfo{volume}{465}}, \bibinfo{pages}{1066} (\bibinfo{year}{2010}).

\bibitem[{\citenamefont{Metzler and Klafter}(2000)}]{metzler}
\bibinfo{author}{\bibfnamefont{R.}~\bibnamefont{Metzler}} \bibnamefont{and}
  \bibinfo{author}{\bibfnamefont{J.}~\bibnamefont{Klafter}},
  \bibinfo{journal}{Physics Reports} \textbf{\bibinfo{volume}{339}},
  \bibinfo{pages}{1} (\bibinfo{year}{2000}).

\bibitem[{\citenamefont{Klages et~al.}(2008)\citenamefont{Klages,
  Radons, and Sokolov}}]{klages}
\bibinfo{author}{\bibfnamefont{R.}~\bibnamefont{Klages}},
  \bibinfo{author}{\bibfnamefont{G.}~\bibnamefont{Radons}}, \bibnamefont{and}
  \bibinfo{author}{\bibfnamefont{I.}~\bibnamefont{Sokolov}},
  \emph{\bibinfo{title}{{Anomalous Transport: Foundations and Applications}}}
  (\bibinfo{publisher}{Wiley-VCH}, \bibinfo{year}{2008}).

\bibitem[{\citenamefont{Heinsalu et~al.}(2010)\citenamefont{Heinsalu,
  Hern\'{a}ndez-Garc\'{\i}a, and L\'{o}pez}}]{Elsref}
\bibinfo{author}{\bibfnamefont{E.}~\bibnamefont{Heinsalu}},
  \bibinfo{author}{\bibfnamefont{E.}~\bibnamefont{Hern\'{a}ndez-Garc\'{\i}a}},
  \bibnamefont{and}
  \bibinfo{author}{\bibfnamefont{C.}~\bibnamefont{L\'{o}pez}},
  \bibinfo{journal}{EPL (Europhysics Letters)} \textbf{\bibinfo{volume}{92}},
  \bibinfo{pages}{40011} (\bibinfo{year}{2010}).

\bibitem[{\citenamefont{Okabe et~al.}(1992)\citenamefont{Okabe, Boots, and
  Sugihara}}]{voronoi}
\bibinfo{author}{\bibfnamefont{A.}~\bibnamefont{Okabe}},
  \bibinfo{author}{\bibfnamefont{B.}~\bibnamefont{Boots}}, \bibnamefont{and}
  \bibinfo{author}{\bibfnamefont{K.}~\bibnamefont{Sugihara}},
  \emph{\bibinfo{title}{{Spatial Tesselations: Concepts and applications of
  Voronoi Diagrams}}} (\bibinfo{publisher}{John Wiley \& Sons},
  \bibinfo{year}{1992}).

\end{thebibliography}
\end{document}